\documentclass[aps, prb]{revtex4-1}
\usepackage[latin9]{inputenc}
\usepackage{bm}
\usepackage{amsmath}
\usepackage{amssymb}
\usepackage{graphicx}

\makeatletter
\usepackage{subfigure}
\usepackage{upgreek}

\makeatother

\begin{document}

\title{Perturbation theories behind thermal mode spectroscopy for high-accuracy
measurement of thermal diffusivity of solids}

\author{Hideshi Ishida}
\email{hideshi.ishida@mec.setsunan.ac.jp}

\affiliation{Department of Mechanical Engineering, Setsunan University\\ 17-8
Ikeda-Nakamachi, Neyagawa, Osaka 572-8508, Japan}

\author{Hirotsugu Ogi}

\affiliation{Department of Precision Science \& Technology, Graduate School of
Engineering, Osaka University\\ 2-1 Yamada-Oka, Suita, Osaka 565-0871,
Japan}
\begin{abstract}
Thermal mode spectroscopy (TMS) has been recently proposed for accurately
measuring thermal diffusivity of solids from a temperature decay rate
of a specific thermal mode selected by three-dimensional (anti)nodal
information {[}Phys. Rev. Lett., \textbf{117}, 195901 (2016){]}. In
this paper, we find out the following advantages of TMS by use of
perturbation analyses. First, TMS is applicable to the measurement
of high thermal diffusivity with a small size specimen. Second, it
is less affected by thermally resistive films on a specimen in the
sense that the resistance at the interface does not affect the first-order
correction of thermal diffusivity. Third, it can perform doubly accurate
measurement of the thermal diffusivity specified at a thermal equilibrium
state even if the diffusivity depends on temperature in the sense
that the measurement can be performed within tiny temperature difference
from the given state and that the decay rate of the slowest decaying
mode is not affected by the dependence.
\end{abstract}
\maketitle

\section{INTRODUCTION\label{sec:intro}}

Thermal conductivity is one of very important properties that characterize
macroscopic state of materials in condensed matter physics. Since
\AA ngstr\"{o}m measured the thermal conductivity of copper et al.
in 1861 \cite{Angstrom1861,*Angstrom1863}, there have been numerous
studies on the measurement of thermal conductivities or diffusivities
of solids. They are categorized into steady, periodic and transient
methods.

The steady (static) methods \cite{steady1880_org,steady73,steady14}
evaluate the thermal conductivity $k$ from steady temperature gradient
for a given constant heat input $Q$. The others are unsteady (dynamic)
methods, and directly or indirectly measure dynamic temperature response
for evaluating thermal diffusivity $a(\equiv k/C)$, where $C$ is
volumetric specific heat: the periodic \cite{Angstrom1861,periodic1915,periodic54}
and transient \cite{transient1924,transient1932,transient06} methods
measure frequency response and short-period overall response of temperature,
respectively. The former includes widely used method such as $3\omega$
methods \cite{3omega87,3omega07}, frequency-domain thermoreflectance
(FDTR) methods \cite{fdtr2009_org}, and photoacoustic methods \cite{photoac76,photoac12}.
The latter includes laser flash methods \cite{laserflush1961_org,laserf01}
and time-domain thermoreflectance (TDTR) methods \cite{tdtr86,tdtr96,tdtr13}.
Almost all other or recent methods \cite{accalory85,twa90_org,phototsg16,FR_Raman16}
are classified into the unsteady method.

In general, however, there are three difficulties in the conventional
methods. First, the measurement of thermal diffusivity for small,
high-thermal-conductivity specimens is difficult. The steady methods
require sufficiently large temperature difference $\Delta T$ in a
specimen to ensure the precision of measured conductivity and, therefore,
a typical length $L$ of specimen over thermal conductivity $k$,
i.e. $L/k$, must be large. For the unsteady methods, the ratio of
typical time scale $L^{2}/a$ of heat conduction to time scale for
the measurement should be sufficiently large. For the periodic methods,
the measurement time scale is the reciprocal of angular frequency
$\omega$ of heat input, and the ratio yields $L^{2}\omega/a$, interpreted
as the squared ratio of the typical length to the thermal diffusion
length $\sqrt{a/\omega}$. For the transient methods, the measurement
time scale must be the time resolution $\delta t$ of measurement
accurately to track the time variation of temperature. Thus, the condition
that $L/k$ must be relatively large is imposed for the conventional
methods, making the measurement of the above-mentioned specimens difficult.

Second, the temperature at which thermal diffusivity is measured can
not be accurately specified. The thermal conductivity depends on both
of temperature and pressure. In general, the pressure over the specimen
can be regarded to be constant for the time scale of thermal diffusion
time. However, the temperature variation in the specimen can not be
ignored. 

As for the steady methods, the above-mentioned temperature difference
$\Delta T$ is involved in the specimen. For the periodic methods,
there exists the amplitude $\Delta T$ of temperature vibration near
a heating surface with the depth of thermal diffusion length, i.e.
the thickness of vibrational thermal boundary layer. The amplitude
$\Delta T$ can be estimated to be the order of $Q/(CA\sqrt{a/\omega})$,
where $A$ is the area of heating spot on the heating surface. Therefore,
the amplitude increases with the frequency. For the transient method,
the temperature difference $\Delta T$ should be caused, like steady
methods, to accurately measure a transient temperature variation.
Thus, to some extent, the temperature variation in a specimen is involved
in the conventional measurement of thermal diffusivity, and the variation
prevents us from identifying the temperature at which the thermal
diffusivity is measured. In this situation, an effective temperature
is required for specifying the state of the measurement. For example,
such a temperature is proposed for the laser flash method \cite{laserflush1961_org}.
Its difference from an initial temperature reaches, however, 1.6 times
that of a final, steady temperature.

Third, the measurement of thermal diffusivity is, in general, affected
by a thin film with thermal resistance on a specimen. In order to
accurately measure temperature variation, we often apply a coating
on a measuring portion of specimen surface, such as melting of solder
or spraying black-body materials. Resultant thin film on the specimen
involves the so-called Kapitza resistance at the film-specimen interface,
which can cause fatal errors for the measurement with small, high-conductivity
specimens.

Recently, we \cite{ogi16} proposed thermal mode spectroscopy (TMS)
to measure the thermal diffusivity of small solids. It is based on
the concept of thermal mode and decay rate corresponding to eigenfunction
and eigenvalue, respectively, defined on the (uniform-diffusivity)
heat conduction equation. The conventional methods observe the propagation
of heat or total temperature variation caused by  a superposition
of thermal modes with the corresponding decay rates. In contrast,
TMS measures the decay rate of a selected thermal mode to obtain the
thermal diffusivity of a specimen. 

It is worthwhile noting that few transient methods can be interpreted
as TMS. Frazier \cite{transient1932} measured the temperature decay
rate of the slowest decaying mode (SDM) excited in a rod to obtain
the thermal diffusivity of nickel with an appropriate selection of
measurement points so that the second slowest mode could be eliminated.
Alwi et al. \cite{tdtr13} applied TDTR to measure the decay rate
of the SDM induced by the heating on the front surface. All the cases
measure the decay rate of the SDM in one-dimensional (axisymmetric
or pointwise symmetric) heat conduction phenomena. These types of
transient method accordingly treat the decay rate of the SDM after
time passes sufficiently. 

It should be emphasized that the method of TMS actively utilizes three-dimensional
(anti)nodal distribution of thermal modes for selectively exciting
and detecting a target mode, not necessarily the SDM, while eliminating
some nearest decay-rate modes. The mode selection is realized by a
 choice of excitation and detection points, executed on a pump-probe
laser system. If an instantaneous heat input is given at a point on
the specimen, some thermal modes which has an antinode at the point
are excited. Among them, the temperature variation of the slowest
decaying mode relative to a detection point (rSDM) is observed whose
antinode is positioned at the detection point. This is the mode-selection
principle \cite{ogi16}.

The method is analogous to the resonant ultrasound spectroscopy \cite{uss93,uss96,uss_ogi99,uss_ogi02},
which is based on vibrational modes and effective for the measurement
of elastic constants of small solids. That is the reason why TMS is
expected to be suitable for the measurement of high thermal diffusivity
of small specimens. We \cite{ogi16} actually succeeded in measuring
high thermal diffusivities of such as diamond et al. with various
millimeter-size specimens. 

In this paper, perturbation analyses are performed to argue that TMS
has advantages to resolve the above-mentioned conventional three problems
and makes it possible to measure thermal diffusivities more accurately.
In the next section, heat conduction and eigenvalue problems are formulated,
and the TMS method is briefly introduced. In the subsequent sections,
the merits of TMS are theoretically and numerically investigated and
discussed.

\section{FUNDAMENTALS OF THERMAL MODE SPECTROSCOPY\label{sec:basic_formulation}}

In case where the thermal diffusivity $a$ is spatially uniform, the
heat conduction equation on the domain $V$ of specimen can be normalized
by the diffusivity, typical length $L$, and typical temperature difference
$\Delta T$, expressed by

\begin{equation}
\frac{\partial\theta}{\partial t}=\triangle\theta,\label{eq:heat_conduction_s}
\end{equation}
where $\triangle$ is the Laplacian.

If the equation has a steady solution under a given boundary condition,
the temperature variation $\theta$ can be divided into the solution
$\theta_{s}$ and vanishing, transient temperature $\theta_{t}$,
i.e. $\theta=\theta_{s}+\theta_{t}$. The transient temperature $\theta_{t}$
is the solution under the zero-valued boundary condition: it is the
so-called adiabatic, free-normal-derivative condition for the Neumann
condition. Substituting an equation $\theta=v(\mathbf{x})exp(-\lambda t)$
into Eq. (\ref{eq:heat_conduction_s}), we obtain

\begin{equation}
-\triangle v=\lambda v,\label{eq:eigenp_thermodes}
\end{equation}
and this is the eigenvalue problem of the Laplace operator. It is
well known that all eigenvalues $\lambda$ are real for the (zero-valued)
Dirichlet, Neumann, Robin boundary conditions and eigenfunctions $v$
can be chosen to be real-valued \cite{walter92}. In addition, eigenfunctions
corresponding to distinct eigenvalues are orthogonal based on the
inner product, defined for the functions $f$ and $g$:

\begin{equation}
(f,\textrm{ }g)\equiv\int_{V}fgdV,\label{eq:org_innerp}
\end{equation}
and the eigenvalues are non-negative for the Neumann condition \cite{walter92}.
The eigenfunction corresponding to the minimum (zero) eigenvalue is
spatially uniform on $V$. 

Since all the eigenvalues are known to be semi-simple, we can prepare
an orthonormal set of eigenfunctions $\hat{v_{i}}\textrm{ }(i=1,\cdots,\infty)$
for expanding the transient temperature $\theta_{t}$ as

\begin{equation}
\theta_{t}(\mathbf{x},t)=\sum_{i=1}^{\infty}a_{i}e^{-\lambda_{i}t}\hat{v_{i}}(\mathbf{x}),\label{eq:theta_t}
\end{equation}
where the eigenvalues $\lambda_{i}$ corresponding to $\hat{v}_{i}$
are arranged in ascending order, $\lambda_{1}\leq\lambda_{2}\leq\cdots$,
and the expansion is based on the completeness of the eigenfunctions
\cite{walter92}. Hereafter, the eigenfunction $v$ or $\hat{v}_{i}$
is called a thermal mode. As described in Eq. (\ref{eq:theta_t}),
a thermal mode $\hat{v}_{i}$ has an intrinsic decay rate $\lambda_{i}$
or a relaxation time $\tau_{i}(\equiv\lambda_{i}^{-1})$, and the
transient temperature variation can be expressed as a superposition
of such thermal modes. Once the $i$th thermal mode is excited and
its dimensional decay rate $\lambda_{i}^{*}$ is measured, the thermal
diffusivity $a^{*}(\equiv k/C)$ of the specimen can be evaluated
to be

\[
a^{*}=\lambda_{i}^{*}L^{2}/\lambda_{i},
\]
and this is the essence of TMS, thermal mode spectroscopy \cite{ogi16}.
In this method, the dimensionless eigenvalue $\lambda_{i}$ must be
known before measurements. As for some simple shapes, such as cuboid,
sphere and circular cylinder, thermal modes and corresponding decay
rates are analytically obtained. In general, it can be computed by
the Rayleigh-Ritz approximation \cite{walter92,ogi16} as discussed
later. 

The thermal modes can be defined on various boundary conditions. Since
we can neither fix the temperature (Dirichlet condition) nor control
the heat transfer rate (Robin condition) accurately on the surface
of specimen because of thermal resistances or ambient fluid flows,
the thermal mode for the adiabatic boundary condition \cite{ogi16}
is particularly important to the measurement of thermal diffusivity.
All we can do is make de-facto insulated condition on the surface,
discussed in the final section. 

The condition, once fulfilled, allows us to measure the thermal diffusivity
without evaluating ambient parameters as needed for most measurement
techniques, simplifying the measurement. Moreover, the temperature
inside a specimen evolves to reach a uniform-temperature, thermal
equilibrium state, and the uniform temperature, at which the diffusivity
is measured, can be accurately specified. They turn out to be advantages
of TMS. That is the reason why thermal modes for the Neumann (adiabatic)
condition are exclusively treated in this paper.

In the next section, we shall see how the thermal mode and the decay
rate are affected by ambiguous parameters.

\section{EFFECTS OF VARIOUS COMPONENTS ON THERMAL MODES - PERTURBATION THEORY
-}

Hereafter we assume that every eigenvalue, formulated in the previous
section, is simple. The condition comes from the fact that TMS utilizes
confirmed antinodal or nodal points of thermal modes to select a target
mode: this is accomplished by the change of excitation and detection
points in measurements. If the eigenvalue corresponding to the target
mode has multiplicity, we cannot know in advance its nodal or antinodal
points. In addition, as the Stark effect, the eigenspace is structurally
unstable, easily divided into one-dimensional eigenspaces by such
minute disturbance as just heat transfer on the surface treated in
this section. At least, the most largest eigenvalues including the
one corresponding to a target mode must be simple. This is the condition
for the application of TMS, easily fulfilled by avoiding highly symmetric
specimens, such as a cube or a regular tetrahedron.

\subsection{Heat transfer on a specimen\label{subsec:effects_h}}

In this section we explore the effects of surface heat transfer rate
on the thermal mode and its decay rate. We can construct a vector
space with the bases of eigenfunction (\ref{eq:eigenp_thermodes})
of the Laplacian. However, it is inapplicable except a specified boundary
condition, i.e. the adiabatic condition in this study: a superposition
of the eigenfunctions on the space does not hold other boundary conditions.
As we shall see, the eigenvectors defined on a finite $N$-dimensional
form of the eigenvalue problem (\ref{eq:eigenp_thermodes}) construct
full $N$-dimensional vector space, yielding a set of bases to investigate
the effects of boundary condition on thermal modes. It is natural,
therefore, to treat a corresponding approximated heat-conduction equation
as a form of sufficiently large $N$-dimensional ordinary differential
equation system (ODEs).

When the heat transfer rate is sufficiently small, it is extremely
difficult numerically to evaluate the effects with high precision:
the smaller the transfer rate, the smaller the effects become. That
is the reason why, in this section, a perturbation problem is raised
and solved analytically. 

\subsubsection{Discretization of heat conduction equation}

The ODEs itself is constructed by the expansion of dimensionless temperature
$\theta$ by use of an appropriate $N$ base functions $\varphi_{\alpha}$
($\alpha=1,\cdots,N$):

\begin{equation}
\theta(\mathbf{x},t)=\sum_{\alpha}\theta_{\alpha}(t)\varphi_{\alpha}(\mathbf{x}).\label{eq:theta_expansion}
\end{equation}
A weak formulation of Eq. (\ref{eq:heat_conduction_s}), or an extension
of the Ritz method \cite{walter92,ogi16}, leads to the following
ODEs with respect to the coefficient $\theta_{\alpha}$ for an arbitrary-shape
domain $V$:

\begin{subequations}\label{eq:unif_goveq_a}

\begin{equation}
\mathbf{M}\frac{d\bm{\uptheta}}{dt}+\mathbf{\Gamma}\bm{\uptheta}+\mathbf{c}=\mathbf{0},\label{eq:goveq_vec_theta}
\end{equation}
where

\begin{equation}
\bm{\uptheta}\equiv(\theta_{1},\cdots,\theta_{N}),\label{eq:def_vec_theta}
\end{equation}

\begin{equation}
\mathbf{M}=\{M_{\alpha\beta}:\textrm{ }\mathit{M_{\alpha\beta}=\int_{V}\varphi_{\alpha}\varphi_{\beta}dV},\textrm{ }1\leq\alpha,\beta\leq N\},\label{eq:M}
\end{equation}

\begin{equation}
\mathbf{\Gamma}=\{\varGamma_{\alpha\beta}:\textrm{ }\mathit{\Gamma_{\alpha\beta}=\int_{V}\sum_{j}\frac{\partial\varphi_{\alpha}}{\partial x_{j}}\frac{\partial\varphi_{\beta}}{\partial x_{j}}dV},\textrm{ }1\leq\alpha,\beta\leq N\},\label{eq:Gamma}
\end{equation}

\begin{equation}
\mathbf{c}=\{c_{\alpha}:\textrm{ }\mathit{c_{\alpha}=\int_{S_{e}}f_{n}\varphi_{\alpha}dS},\textrm{ }1\leq\alpha\leq N\}.\label{eq:c_alpha}
\end{equation}
\end{subequations}Herein, the surface $S_{e}$ is subjected to a
boundary condition except adiabatic condition, and $f_{n}$ denotes
the heat flux outward normal to the surface at a boundary point. As
defined in Eqs. (\ref{eq:M}) and (\ref{eq:Gamma}), $\mathbf{M}$
and $\mathbf{\Gamma}$ are positive-definite and positive-semidefinite
symmetric matrices, respectively. The semidefiniteness comes from
the fact that Eq. (\ref{eq:goveq_vec_theta}) has the steady (constant)
solution for the adiabatic condition, i.e. when $\mathbf{c}=\mathbf{0}$. 

The flux $f_{n}$ is independent of the temperature $\bm{\uptheta}$
under the Neumann (adiabatic) condition. Then substitution of $\bm{\uptheta}=e^{-\hat{\lambda}t}\hat{\mathbf{v}}$
into Eq. (\ref{eq:goveq_vec_theta}) produces the following eigenvalue
problem (Rayleigh-Ritz approximation \cite{walter92,ogi16}):

\begin{equation}
\mathbf{\Gamma\hat{v}}_{i}=\hat{\lambda}_{i}\mathbf{M\hat{v}}_{i},\textrm{ }(i=1,\cdots,N).\label{eq:fund_eigenv_problem}
\end{equation}

The above-mentioned property shows that the eigenvalue $\hat{\lambda}$
is non-negative real number and the two eigenvectors $\mathbf{\hat{v}}_{i}$
and $\mathbf{\hat{v}}_{j}$ corresponding to different eigenvalues
$\hat{\lambda}_{i}$ and $\hat{\lambda}_{j}$, respectively, are orthogonal
in the sense that $(\mathbf{\hat{v}}_{i},\mathbf{\hat{v}}_{j})_{M}\equiv\mathbf{\hat{v}}_{i}^{T}\mathbf{M}\mathbf{\hat{v}}_{j}=0$;
the inner product corresponds to Eq. (\ref{eq:org_innerp}). These
properties reflect those of the original eigenvalue problem (\ref{eq:eigenp_thermodes}),
and the vector $\mathbf{\hat{v}}_{i}$ is a discretized expression
of each thermal mode under the Neumann condition, $\hat{v_{i}}$.
Assuming all eigenvalues are simple, they can be aligned to hold $0=\hat{\lambda}_{1}<\hat{\lambda}_{2}<\cdots<\hat{\lambda}_{N}$,
and the eigenvectors can be orthonormalized, i.e. $(\mathbf{\hat{v}}_{i},\mathbf{\hat{v}}_{j})_{M}=\delta_{ij}$.
Thus the union of all of the eigenspaces forms the full $N$-dimensional
vector space, allowing us to discuss the change in eigenvalues and
eigenfunctions caused by a (slight) change of boundary condition on
the same eigenbases $\mathbf{\hat{v}}_{i}\textrm{ }(i=1,\cdots,N)$.

The selection of base function is important in this procedure. They
should not hold a specified boundary condition, and the eigenvalue
$\hat{\lambda}_{i}$ and eigenvector $\mathbf{\hat{v}}_{i}$ should
converge to the $i$th eigenvalue and eigenfunction, respectively,
of the problem (\ref{eq:eigenp_thermodes}) as $N\rightarrow\infty$.
For the Neumann condition, it is known \cite{walter92} that any $C^{2}$
non-zero functions $\varphi_{\alpha}$ ($\alpha=1,\cdots,N$) on $V$
are appropriate for the Rayleigh-Ritz approximation (\ref{eq:fund_eigenv_problem})
in the sense that the roots of the equation are approximation to the
first $N$ eigenvalues of the problem (\ref{eq:eigenp_thermodes}).
It should be noted, however, that the choice does not affect the form
(\ref{eq:unif_goveq_a}) and then the following formal discussions.
Therefore, we only assume that the first base function $\varphi_{1}$
is constant at $1/\sqrt{V}$. The choice gives $\varGamma_{1\alpha}=0$,
permitting a constant-temperature steady solution $\bm{\uptheta}_{s}$
under the adiabatic condition to be $\mathbf{\uptheta}_{s}=(\theta_{s},0,\cdots,0)$.
This is the first eigenvector $\mathbf{\hat{v}}_{1}$ corresponding
to the minimum eigenvalue $\hat{\lambda}_{1}(=0)$. 

\subsubsection{Effects of heat transfer on a specimen surface}

If a constant heat-transfer rate $h$ is given on a portion $S_{e}$
in a specimen surface, the non-dimensional evolution equation (\ref{eq:goveq_vec_theta})
of temperature has the vector $\mathbf{c}$ (\ref{eq:c_alpha}) of
the form

\[
\mathbf{c}=\{c_{\alpha}:\textrm{ }\mathit{c_{\alpha}=Bi\int_{S_{e}}(\sum_{\beta}\theta_{\beta}\varphi_{\beta}-\theta_{\infty})\varphi_{\alpha}dS},\textrm{ }1\leq\alpha\leq N\},
\]
where $Bi$ denotes the Biot number, defined by $Bi\equiv hL/k$,
$\theta_{\infty}$ ambient temperature outside the specimen $V$.
The boundary condition is the so-called Robin type. The vector poses
an eigenvalue problem

\begin{equation}
\mathbf{(\Gamma+\mathit{Bi}\Gamma')v}_{i}=\lambda_{i}\mathbf{Mv}_{i},\textrm{ }(i=1,\cdots,N),\label{eq:eigenp_Bi}
\end{equation}
where $\mathbf{\Gamma}'$ is a positive-definite symmetric matrix,
given by

\[
\mathbf{\Gamma'}=\{\varGamma'_{\alpha\beta}:\textrm{ }\mathit{\Gamma'_{\alpha\beta}=\int_{S_{e}}\varphi_{\alpha}\varphi_{\beta}dS},\textrm{ }1\leq\alpha,\beta\leq N\}.
\]

While the Biot number is sufficiently small, we can use the number
as a perturbation parameter $\epsilon$ and present a perturbation
theory. Expanding the variables as follows \begin{subequations}\label{eq:expand_vars}

\begin{equation}
\lambda_{i}=\hat{\lambda_{i}}+\epsilon\lambda_{i}^{(1)}+\epsilon^{2}\lambda_{i}^{(2)}+\cdots,\label{eq:expand_lambda}
\end{equation}

\begin{equation}
\mathbf{v}_{i}=\mathbf{\hat{v}}_{i}+\epsilon\mathbf{v}_{i}^{(1)}+\epsilon^{2}\mathbf{v}_{i}^{(2)}+\cdots,\label{eq:expand_x}
\end{equation}

\begin{equation}
\mathbf{v}_{i}^{(\alpha)}=\sum_{k}w_{i,k}^{(\alpha)}\mathbf{\hat{v}}_{k},\label{eq:expand_xalpha}
\end{equation}

\end{subequations}\noindent and substituting into Eq. (\ref{eq:eigenp_Bi})
we obtain up to the second-order corrections\begin{subequations}\label{eq:Biot_results}

\begin{equation}
\lambda_{i}^{(1)}=\mathbf{\hat{v}}_{i}^{T}\Gamma'\mathbf{\hat{v}}_{i},\label{eq:lamd_sup1}
\end{equation}

\begin{equation}
w_{i,k}^{(1)}=\frac{\mathbf{\hat{v}}_{k}^{T}\Gamma'\mathbf{\hat{v}}_{i}}{\hat{\lambda_{i}}-\hat{\lambda_{k}}},\textrm{ }(k\neq i),\label{eq:C_i,k^(1)_ht}
\end{equation}

\begin{equation}
w_{i,i}^{(1)}=0,\label{eq:C_ii^(1)_ht}
\end{equation}

\begin{equation}
\lambda_{i}^{(2)}=\mathbf{\hat{v}}_{i}^{T}\Gamma'\mathbf{v}_{i}^{(1)},\label{eq:lam_i^(2)_ht}
\end{equation}

\begin{equation}
w_{i,k}^{(2)}=\frac{-\lambda_{i}^{(1)}w_{i,k}^{(1)}+\mathbf{\hat{v}}_{k}^{T}\Gamma'\mathbf{v}_{i}^{(1)}}{\hat{\lambda_{i}}-\hat{\lambda_{k}}},\textrm{ }(k\neq i),\label{eq:C_ik^(2)_ht}
\end{equation}

\begin{equation}
w_{i,i}^{(2)}=-\frac{1}{2}\sum_{k}w_{i,k}^{(1)^{2}}.\label{eq:C_ii^(2)_ht}
\end{equation}

\end{subequations}

\noindent Herein the coefficient $w_{i,i}^{(\cdot)}$ is obtained
from the normalization condition of $\mathbf{v}_{i}$. 

For instance, Eq. (\ref{eq:lamd_sup1}) allows us to evaluate the
first-order correction $\lambda_{1}^{(1)}$ of the eigenvalue $\hat{\lambda}_{1}$
as a non-dimensional shape factor $S_{e}/V$. Once the shape of specimen
is fixed, however, we cannot make these corrections in Eq. (\ref{eq:Biot_results})
sufficiently small. We can instead reduce $Bi$ in order for the application
of adiabatic (Neumann-type) thermal modes as discussed later. Conversely,
the conventional methods need for $L/k$ to be sufficiently large,
as discussed in Sec. \ref{sec:intro} and, as a result, $Bi$ must
be large. The Biot number is, therefore, the most important dimensionless
variable to quantify the precision of TMS, and also to characterize
TMS among conventional techniques.

\subsection{Thin films with thermal resistance on a specimen\label{subsec:effects_resistive-films}}

As explained in Sec. \ref{sec:intro}, most measurement techniques
involve thin films with the Kapitza resistance on a specimen. Hereafter,
the films are referred to as the Kapitza resistive films. TMS also
requires such a film particularly for a transparent (translucent)
specimen, and it is important to investigate the effects of a film
on both of thermal modes and decay rates. Similar to the previous
section, it is difficult for us to treat numerically the effects with
high precision when the film is very thin: spatial and temporal resolution
is determined by the thin film, and the numbers of base functions
and time steps become larger as the thickness of the film decreases.
On the platform of the previously described ODEs, however, we can
derive a heat conduction equation on an arbitrary-shape specimen with
some resistive films, and raise another perturbation problem for quantifying
the effects. 

The derivation of the equation is detailed in Appendix \ref{sec:hceq_Kap}.
In this section, we consider a single resistive film because all important
properties of thermal modes are obtained from the equation (\ref{eq:finaleq_theta})
for the case that the number $R$ of resistive films equals one, and
qualitative discussions are unchanged by the multiple films. 

From the evolution equation, the problem to obtain thermal modes and
their decay rates is reduced to the following eigenvalue problem:

\begin{equation}
(\mathbf{\Gamma}+\mathbf{N}\mathbf{A}'_{0})\mathbf{v}_{i}=\left[\lambda_{i}\mathbf{M}-\mathbf{N}(\lambda_{i}\mathbf{A}'_{1}+\lambda_{i}^{2}\mathbf{A}'_{2}+\cdots)\right]\mathbf{v}_{i},\textrm{ }(i=1,\cdots,N),\label{eq:final_eigenp}
\end{equation}
and we are to solve the equation by another perturbation analysis. 

If the (typical) thickness $l$ of the resistive film, normal to the
specimen surface, is sufficiently small, we can use $l$ as a perturbation
parameter $\epsilon$ and expand the quantities (\ref{eq:goveq_rfilm})
relating to the film as follows \begin{subequations}

\begin{align}
\mathbf{M'} & =\epsilon\mathbf{M}_{1}+\epsilon^{2}\mathbf{M}_{2}+\cdots=\epsilon C'\mathbf{N}+\epsilon^{2}\mathbf{M}_{2}+\cdots,\label{eq:M'_exp}\\
\mathbf{\Gamma}' & =\epsilon\mathbf{\Gamma}_{1}+\epsilon^{2}\mathbf{\Gamma}_{2}+\cdots=\epsilon k'\mathbf{D}+\epsilon^{2}\mathbf{\Gamma}_{2}+\cdots,\label{eq:Gam'_exp}\\
\mathbf{A}'_{\alpha} & =\epsilon^{\alpha}\mathbf{A}{}_{\alpha\alpha}+\epsilon^{\alpha+1}\mathbf{A}{}_{\alpha,\alpha+1}+\cdots,\label{eq:A'_alpha}
\end{align}

\end{subequations}\noindent where

\[
\mathbf{D}=\{D{}_{\alpha\beta}:\textrm{ }\mathit{D_{\alpha\beta}=\int_{Se}\sum_{j}\frac{\partial\varphi_{\alpha}}{\partial x_{j}}\frac{\partial\varphi_{\beta}}{\partial x_{j}}dS},\textrm{ }1\leq\alpha,\beta\leq N\},
\]
and other quantities except the corrections are defined in Eqs. (\ref{eq:goveq_rfilm})
and (\ref{eq:def_N}). Herein, the fact that $\mathbf{A}'_{\alpha}$
is the order of $O(l^{\alpha})$ is reflected on Eq. (\ref{eq:A'_alpha}).
In addition, we again expand the eigenvalue and eigenvector of the
form (\ref{eq:expand_vars}), based on the assumption that all eigenvalues
are simple.

From the definitions of $\mathbf{A}'$ and $\mathbf{A}'_{\alpha}$
(Eq. (\ref{eq:def_As})), it is straightforward to obtain

\begin{align}
\mathbf{A}_{\alpha\beta} & =\frac{(-1)^{\alpha}}{\alpha!\beta!}\left.\frac{\partial^{\alpha+\beta}\mathbf{A}'}{\partial s^{\alpha}\partial l^{\beta}}\right|_{l=s=0}\nonumber \\
 & =(-1)^{\alpha}\sum_{i=max(\alpha,1)}^{\beta}(-l_{K})^{i-1}\sum_{C_{i}^{\alpha,\beta}(n_{j},m_{j})}\prod_{j=1}^{i}(\mathbf{N}^{-1}\mathbf{M}_{m_{j}})^{n_{j}}(\mathbf{N}^{-1}\mathbf{\Gamma}_{m_{j}})^{1-n_{j}},\label{eq:A_alpbet}
\end{align}
where $C_{i}^{\alpha,\beta}(n_{j},m_{j})$ indicates the possible
combinations of integers $n_{j}$ and $m_{j}$, such that

\[
\sum_{j=1}^{i}n_{j}=\text{\ensuremath{\alpha}},\textrm{ }n_{j}=0\textrm{ or }1,
\]
and

\[
\sum_{j=1}^{i}m_{j}=\beta,\textrm{ }m_{j}\geq1.
\]

The equation (\ref{eq:A_alpbet}) makes it possible for us to have
lower-order coefficients of $\mathbf{A}'$ as follows:\begin{subequations}\label{eq:coefs_A'}

\begin{align}
\mathbf{A}_{00} & =\mathbf{0},\label{eq:A_00}\\
\mathbf{A}_{01} & =k'\mathbf{N}^{-1}\mathbf{D},\label{eq:A_01}\\
\mathbf{A}_{02} & =\mathbf{N}^{-1}\mathbf{\Gamma}_{2}-l_{K}k'^{2}(\mathbf{N}^{-1}\mathbf{D})^{2},\label{eq:A_02}\\
\mathbf{A}_{11} & =-C'\mathbf{I},\label{eq:A_11}\\
\mathbf{A}_{12} & =-\mathbf{N}^{-1}\mathbf{M}_{2}+2l_{K}k'C'\mathbf{N}^{-1}\mathbf{D},\label{eq:A_12}\\
\mathbf{A}_{22} & =-C'^{2}l_{K}\mathbf{I}.\label{eq:A_22}
\end{align}

\end{subequations}Substituting Eqs. (\ref{eq:A'_alpha}) and (\ref{eq:coefs_A'})
into Eq. (\ref{eq:final_eigenp}), we have the first- and second-order
equations from which we can obtain \begin{subequations}\label{eq:cors_temp_depend}

\begin{align}
\lambda_{i}^{(1)} & =k'\mathbf{\hat{v}}_{i}^{T}\mathbf{D}\mathbf{\hat{v}}_{i}-C'\hat{\lambda_{i}}\mathbf{\hat{v}}_{i}^{T}\mathbf{N}\mathbf{\hat{v}}_{i},\label{eq:lamd(1)_tempd}\\
w_{i,k}^{(1)} & =\frac{k'\mathbf{\hat{v}}_{k}^{T}\mathbf{D}\mathbf{\hat{v}}_{i}-C'\hat{\lambda_{i}}\mathbf{\hat{v}}_{k}^{T}\mathbf{N}\mathbf{\hat{v}}_{i}}{\hat{\lambda_{i}}-\hat{\lambda_{k}}},\textrm{ }(k\neq i),\label{eq:c_i,k^(1)_tempd}\\
w_{i,i}^{(1)} & =0,\label{eq:c_ii^(1)_tempd}\\
\lambda_{i}^{(2)} & =\mathbf{\hat{v}}_{i}^{T}\mathbf{N}\mathbf{A}_{01}\mathbf{v}_{i}^{(1)}+\mathbf{\hat{v}}_{i}^{T}\mathbf{N}\mathbf{A}_{02}\mathbf{\hat{v}}_{i}-\lambda_{i}^{(1)}(\mathbf{\hat{v}}_{i},\mathbf{v}_{i}^{(1)})_{M}+\hat{\lambda_{i}}\mathbf{\hat{v}}_{i}^{T}\mathbf{N}\mathbf{A}_{11}\mathbf{v}_{i}^{(1)}\nonumber \\
 & +\lambda_{i}^{(1)}\mathbf{\hat{v}}_{i}^{T}\mathbf{N}\mathbf{A}_{11}\mathbf{\hat{v}}_{i}+\hat{\lambda_{i}}\mathbf{\hat{v}}_{i}^{T}\mathbf{N}\mathbf{A}_{12}\mathbf{\hat{v}}_{i}+\hat{\lambda_{i}}^{2}\mathbf{\hat{v}}_{i}^{T}\mathbf{N}\mathbf{A}_{22}\mathbf{\hat{v}}_{i},\label{eq:lamd(2)_tempd}\\
w_{i,k}^{(2)} & =\frac{1}{\hat{\lambda_{i}}-\hat{\lambda_{k}}}\left(\mathbf{\hat{v}}_{k}^{T}\mathbf{N}\mathbf{A}_{01}\mathbf{v}_{i}^{(1)}+\mathbf{\hat{v}}_{k}^{T}\mathbf{N}\mathbf{A}_{02}\mathbf{\hat{v}}_{i}-\lambda_{i}^{(1)}(\mathbf{\hat{v}}_{k},\mathbf{v}_{i}^{(1)})_{M}+\hat{\lambda_{i}}\mathbf{\hat{v}}_{k}^{T}\mathbf{N}\mathbf{A}_{11}\mathbf{v}_{i}^{(1)}\right.\nonumber \\
 & \left.+\lambda_{i}^{(1)}\mathbf{\hat{v}}_{k}^{T}\mathbf{N}\mathbf{A}_{11}\mathbf{\hat{v}}_{i}+\hat{\lambda_{i}}\mathbf{\hat{v}}_{k}^{T}\mathbf{N}\mathbf{A}_{12}\mathbf{\hat{v}}_{i}+\hat{\lambda_{i}}^{2}\mathbf{\hat{v}}_{k}^{T}\mathbf{N}\mathbf{A}_{22}\mathbf{\hat{v}}_{i}\right),\textrm{ }(k\neq i),\label{eq:c_i,k^(2)_tempd}\\
w_{i,i}^{(2)} & =-\frac{1}{2}\sum_{k}w_{i,k}^{(1)^{2}}.\label{eq:c_i,i^(2)_tempd}
\end{align}

\end{subequations}

We can find that the first-order corrections are independent of the
Kapitza conductance, or the normalized Kapitza length $l_{K}$. As
the second-order coefficients $\mathbf{\Gamma}_{2}$ and $\mathbf{M}_{2}$,
appeared respectively in $\mathbf{A}_{02}$ (\ref{eq:A_02}) and $\mathbf{A}_{12}$
(\ref{eq:A_12}), have information on the shape of film domain, $V'$,
we can also find that the second-order corrections are affected by
the shape of films in addition to the conductance. 

\subsection{Temperature dependence of thermal conductivity\label{subsec:temp-dependency}}

Most measurements of thermal diffusivity necessarily introduce measurable
temperature variation within a specimen in the scale of either thermal
diffusion length (periodic method) or whole length (steady or transient
method). The variation causes errors in evaluated thermal diffusivities
through temperature-dependent thermal diffusivity.

As explained in Sec. \ref{sec:intro}, some transient methods, including
the present TMS, measure a temperature decay rate or its relating
variables near equilibrium state. In this section, a perturbation
analysis is conducted to precisely examine the effects of the temperature
dependence on the decay rate, followed by numerical verifications.

\subsubsection{Perturbation analysis on the effects of temperature dependence on
thermal modes}

Hereafter, we assume that the dimensionless thermal diffusivity $a$,
normalized by typical thermal diffusivity, depends only on temperature
and that the temperature evolves under the adiabatic boundary condition.
If $a$ is spatial constant of unity, thermal modes (eigenfunctions)
for the Neumann condition can be defined so that every temperature
field can be expanded by a set of orthonormalized eigenfunctions,
based on their completeness explained in Sec. \ref{sec:basic_formulation}.

Let us begin with the heat conduction equation on a specimen domain
$V$:

\begin{equation}
\frac{\partial\theta}{\partial t}=\sum_{j}\frac{\partial}{\partial x_{j}}\left(a(\theta)\frac{\partial\theta}{\partial x_{j}}\right).\label{eq:tempd_heateq}
\end{equation}
Herein, the temperature $\theta$ is normalized as $\theta=(T-T_{e})/\Delta T$,
where $T$ denotes dimensional temperature field, $T_{e}$ constant
temperature at equilibrium, $\Delta T$ typical temperature difference.

We are to solve the above nonlinear equation around the steady (equilibrium)
state by another perturbation analysis. Around the steady temperature
($\theta=0$), the thermal diffusivity can be expanded by\begin{subequations}\label{eq:expand_tempd}

\begin{align}
a(\epsilon\theta) & =a(0)+\left.\frac{\partial a}{\partial\theta}\right|_{\theta=0}\epsilon\theta+\frac{1}{2!}\left.\frac{\partial^{2}a}{\partial\theta^{2}}\right|_{\theta=0}(\epsilon\theta)^{2}+\cdots,\nonumber \\
 & \equiv1+a_{1}\epsilon\theta+a_{2}\epsilon^{2}\theta^{2}+\cdots,\label{eq:expand_a}\\
\theta & =\theta_{0}+\epsilon\theta_{1}+\epsilon^{2}\theta_{2}+\cdots,\label{eq:expand_theta}
\end{align}
\end{subequations}\noindent where the perturbation parameter $\epsilon(=1)$
is not more than an indicator that represents the order of expansion.
Please note that the order $n$ is combined with $a_{n}$ and, therefore,
the effects of $a_{n}$ first appear in the $n$th-order correction
$\text{\ensuremath{\theta}}_{n}$ of temperature.

Substituting Eq. (\ref{eq:expand_tempd}) into Eq. (\ref{eq:tempd_heateq}),
and after some algebras we obtain the $n$th order equation as follows

\begin{equation}
\frac{\partial\theta_{n}}{\partial t}=\triangle(\theta_{n}+f_{n}),\label{eq:order_n_temp}
\end{equation}
where

\[
f_{n}=\sum_{j=1}^{n}\frac{a_{j}}{j+1}S_{n-j}^{(j+1)},
\]

\[
S_{l}^{(m)}=\sum_{C_{l}^{(m)}(n_{j})}m!\prod_{j=0}^{\infty}\frac{\theta_{j}^{n_{j}}}{n_{j}!},
\]
and the condition $C_{l}^{(m)}(n_{j})$ indicates the possible combinations
of non-negative integers $n_{j}\textrm{ }(j=0,\cdots,\infty)$, such
that

\[
\sum_{j=0}^{\infty}n_{j}=m,\textrm{ and }\sum_{j=0}^{\infty}jn_{j}=l.\textrm{ }
\]

If $\theta_{i}\textrm{ }(i<n)$ satisfies the adiabatic condition
on the surface $S$ of the domain $V$, we can easily confirm that

\[
\int_{V}\triangle f_{n}dV=0.
\]
It follows that both of the steady and transient components of the
$n$th-order temperature $\theta_{n}$ exists \cite{walter92}.

If we set initial conditions as 

\[
\theta_{n}(\mathbf{x},0)=\begin{cases}
\theta(\mathbf{x},0); & n=0,\\
0; & n>0,
\end{cases}
\]
then the steady component of $\theta_{n}$ vanishes for $n>0$.

The equation for $\theta_{0}$ is just the conventional heat conduction
equation (\ref{eq:heat_conduction_s}) and, therefore, its transient
part $\theta_{0}^{(t)}$ can be expanded by the form (\ref{eq:theta_t})
with eigenfunctions (thermal modes) $\hat{v_{i}}\textrm{ }(i=1,\cdots,\infty)$
for the Neumann condition. This is also the case for higher-order
temperatures $\theta_{i}\textrm{ }(i>0)$. In what follows, as Secs.
\ref{subsec:effects_h} and \ref{subsec:effects_resistive-films},
we assume that every eigenvalue $\lambda_{i}$ is simple.

Substituting the following expansions

\begin{align*}
\theta_{0}^{(t)}(\mathbf{x},t) & =\sum_{i=1}^{\infty}\hat{c}_{i}e^{-\lambda_{i}t}\hat{v_{i}}(\mathbf{x}),\\
\theta_{1}(\mathbf{x},t) & =\sum_{i=1}^{\infty}c_{i}^{(1)}(t)\hat{v_{i}}(\mathbf{x}),
\end{align*}
into the first-order equation (Eq. (\ref{eq:order_n_temp}), $n=1$)
and solving the equation for $c_{i}^{(1)}$, we obtain

\begin{equation}
c_{i}^{(1)}(t)=-\sum_{j,k}\frac{a_{1}\lambda_{i}}{2}\hat{c}_{j}\hat{c}_{k}(\hat{v}_{j}\hat{v}_{k},\hat{v}_{i})\theta_{j,k}^{(i)}(t),\label{eq:c_i^(1)}
\end{equation}
where

\[
\theta_{j,k}^{(i)}(t)=\begin{cases}
te^{-\lambda_{i}t}; & \lambda_{j}+\lambda_{k}-\lambda_{i}=0,\\
\frac{e^{-\lambda_{i}t}-e^{-(\lambda_{j}+\lambda_{k})t}}{\lambda_{j}+\lambda_{k}-\lambda_{i}}; & \lambda_{j}+\lambda_{k}-\lambda_{i}\neq0,
\end{cases}
\]
and $(\cdot,\cdot)$ denotes the inner product defined by Eq. (\ref{eq:org_innerp}).

The coefficient $c_{i}^{(1)}$ shows the first-order correction to
the decaying component of the $i$th eigenfunction $\hat{v}_{i}$
when $a_{1}\neq0$. Note that it does exist even if the dependency-free
component $\hat{c}_{i}$ equals zero as long as $(\hat{v}_{j}\hat{v}_{k},\hat{v}_{i})$
is not zero.

When time $t$ is sufficiently large, $\lambda_{i}t$ is far greater
than $\ln t$. It follows that both of the terms proportional to $te^{-\lambda_{i}t}$
and $e^{-\lambda_{i}t}$ are observed as the ones whose decay rates
are $\lambda_{i}$, identical with the rate of the $i$th thermal
mode. Such a correction, therefore, is unable to be distinguished
from the decay rate for the case of constant diffusivity.

In contrast, the terms expressed by $e^{-(\lambda_{j}+\lambda_{k})t}$
have distinguishable decay rates. If the slowest decaying mode (SDM)
$v$ corresponding to the eigenvalue $\lambda$ dominates after time
passes sufficiently, such terms are observed in total as ones of the
decay rate $2\lambda$ on the $i$th mode when the square $v^{2}$
of the mode is not orthogonal to $\hat{v}_{i}$. If the $i$th, target
mode is just the SDM, the decay rate is twice the target rate, separable
in measurement. 

It is surprising that the temperature dependency does not modify the
decay rate of SDM, as discussed in Secs. \ref{subsec:effects_h} and
\ref{subsec:effects_resistive-films}. In this sense, the SDM feels
its decay rate at thermal equilibrium even in the transient state
of temperature relaxation, not feeling a continuous, time-dependent
change of the rate; that is the notable property of TMS for the accurate
measurement of the thermal diffusivity.

Similarly, the equation for the $n$th-order temperature $\theta_{n}$
can be solved. However, such a correction, as the whole, has larger
decay rate, typically $(n+1)\lambda$, and decays too rapidly to be
observed. As a result, the parameter $a_{1}$, including in Eq. (\ref{eq:c_i^(1)}),
must be the most important dimensionless variable to quantify the
effects of temperature-dependent thermal diffusivity. This is numerically
confirmed in Appendix \ref{sec:depnd_ind}.

\subsubsection{Numerical simulations on the effects of temperature-dependent thermal
diffusivity\label{subsec:simulation}}

\begin{figure}
\includegraphics[scale=0.4]{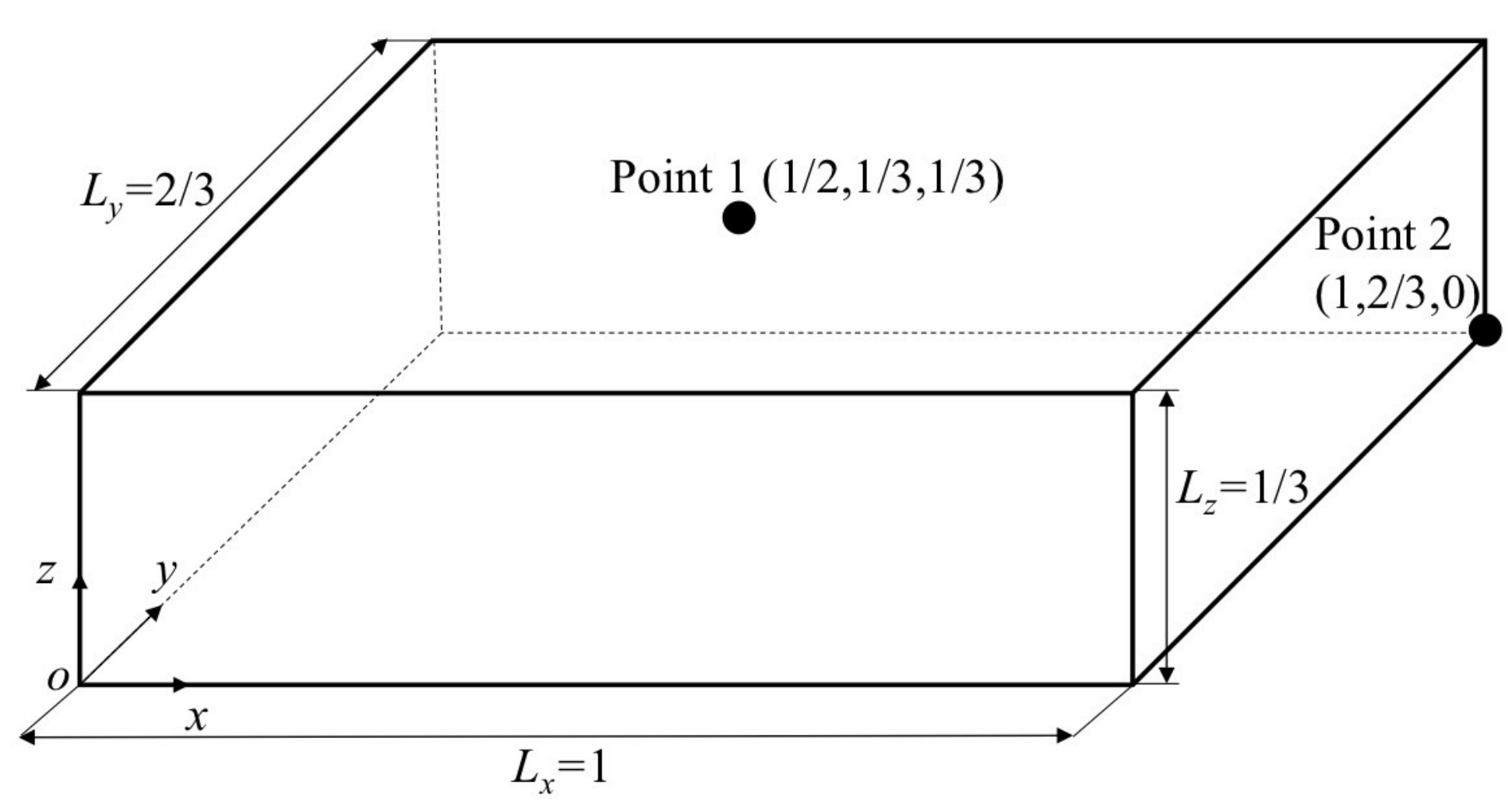}\caption{Physical model and coordinate definition. Two detection points 1 and
2 are set on the surface.\label{fig1}}
\end{figure}

Hereafter, we shall address the heat conduction problem on the cuboid,
denoted by domain $V$, as shown in Fig. \ref{fig1}. Its width $L_{x}$,
depth $L_{y}$, and height $L_{z}$ are 1, 2/3 and 1/3, respectively.
The surface of the cuboid is completely insulated. Two temperature
detection points No. 1 and 2 are set at (1/2,1/3,1/3), and (1,2/3,0),
respectively.

Initially the dimensional temperature $T$ in the cuboid is uniform
at $T_{0}\textrm{ }(\theta=-1)$. At $t=0$, a uniform heat input
$Q/(L_{y}L_{z})$ is given on the surface at $x=0$, introducing a
temperature rise $\Delta T[=Q/(CV)]$ in the steady (equilibrium)
state through the instantaneous total heat input $Q$ so that the
steady temperature $T_{e}=T_{0}+\Delta T$. The rise provides the
typical temperature difference to normalize the dimensional temperature
$T$ as explained in the previous section. 

In this section, a temperature-dependent thermal diffusivity of the
form

\[
a(\theta)=\left(\frac{c_{2}}{\theta+c_{2}}\right)^{c_{1}};\textrm{ power-law type}\textrm{ }(a_{1}=-c_{1}/c_{2}),
\]
is considered, where coefficients $c_{i}\textrm{ }(i=1,2)$ are supposed
to be positive number: $c_{1}$ is changed from 0 to 2.0, and $c_{2}$
is fixed at 3001 or 30001. The range of the parameters is typical
for most pure metals (see Appendix \ref{sec:depnd_ind}).

\begin{figure}
\includegraphics[scale=0.3]{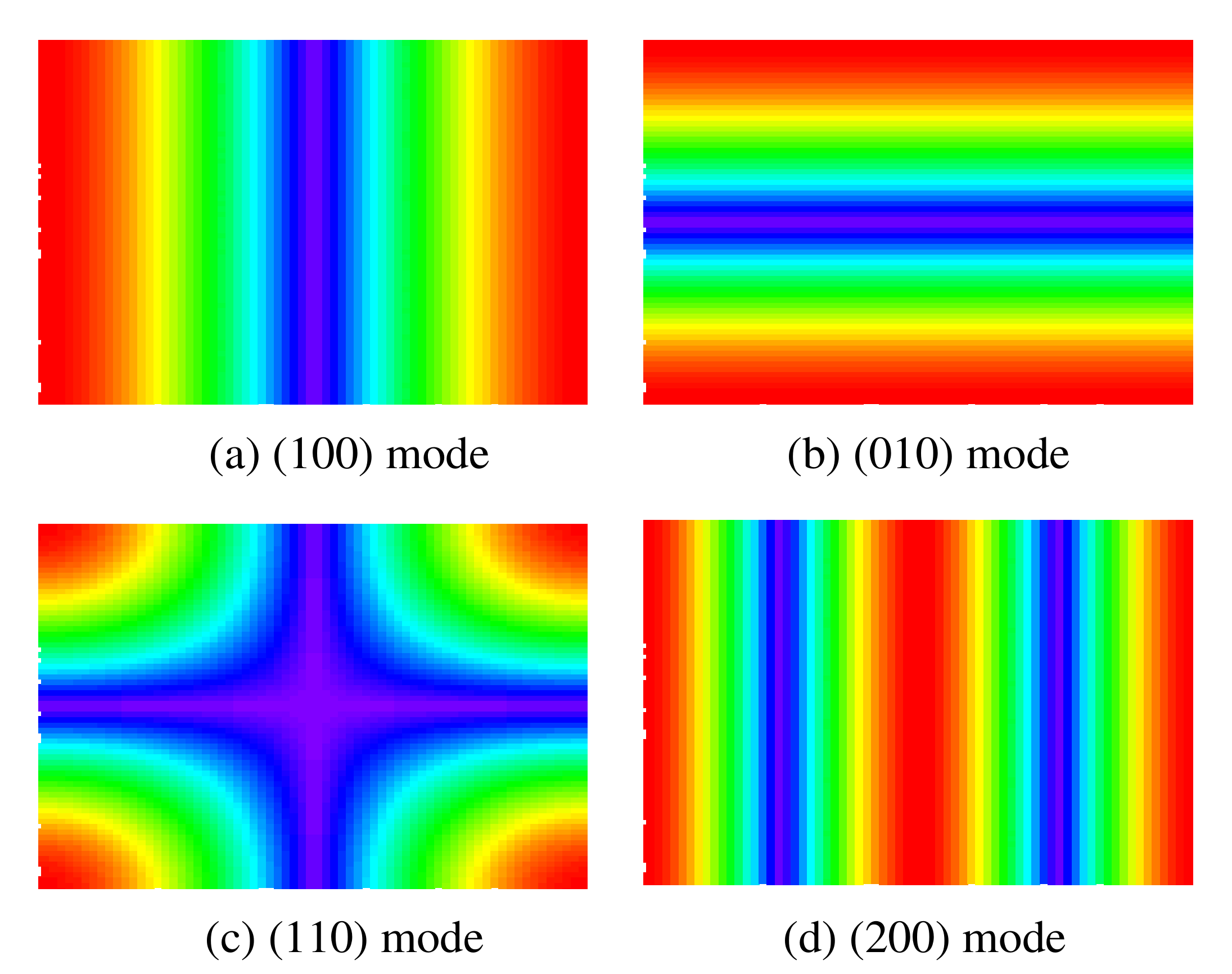}

\caption{Top view of the first four slowest decaying modes ($z=1/3$): (a)
(100) mode; (b) (010) mode; (c) (110) mode; (d) (200) mode. These
figures show the absolute value of $\hat{v}_{lmn}$: the antinodal
($\left|\hat{v}\right|\fallingdotseq3\sqrt{\psi(l)\psi(m)/2}$) and
nodal ($\left|\hat{v}\right|\fallingdotseq0$) regions are drawn in
red and blue, respectively.\label{fig2}}
\end{figure}

The heat conduction equation (\ref{eq:tempd_heateq}) was numerically
solved by the finite volume method (FVM \cite{smith65,patanker}).
The diffusion term was discretized by the central difference scheme,
and time integration was conducted by the first-order Euler explicit
method. A used grid was $151\times101\times50$, and a time step was
$3.66\times10^{-6}$. The resolution can be numerically verified by
comparison with the results on the double-resolution grid. The computations
were performed by double precision.

The cuboid has the eigenfunction $\hat{v}_{lmn}$ and eigenvalue $\lambda_{lmn}$,
specified by three non-negative integers $(lmn)$ as follows \cite{walter92,ogi16}:
\begin{align*}
\lambda_{lmn} & =\pi^{2}\left[\left(\frac{l}{L_{x}}\right)^{2}+\left(\frac{m}{L_{y}}\right)^{2}+\left(\frac{n}{L_{z}}\right)^{2}\right],\\
\hat{v}_{lmn}(\mathbf{x}) & =\sqrt{\frac{\psi(l)\psi(m)\psi(n)}{L_{x}L_{y}L_{Z}}}\cos\left(\frac{l\pi x}{L_{x}}\right)\cos\left(\frac{m\pi y}{L_{y}}\right)\cos\left(\frac{n\pi z}{L_{z}}\right),
\end{align*}
where $\psi(x)\equiv\min(2,x+1)$. Then the non-zero smallest four
eigenvalues of the present model are

\[
\lambda_{100}\fallingdotseq9.86960,\textrm{ }\lambda_{010}\fallingdotseq22.2066,\textrm{ }\lambda_{110}\fallingdotseq32.0762,\textrm{ }\lambda_{200}\fallingdotseq39.4784,
\]
and, therefore, corresponding eigenfunctions are the four most slowest
decaying modes. Their nodal and antinodal regions are illustrated
in Fig. \ref{fig2}.

\begin{figure}
\includegraphics[scale=0.7]{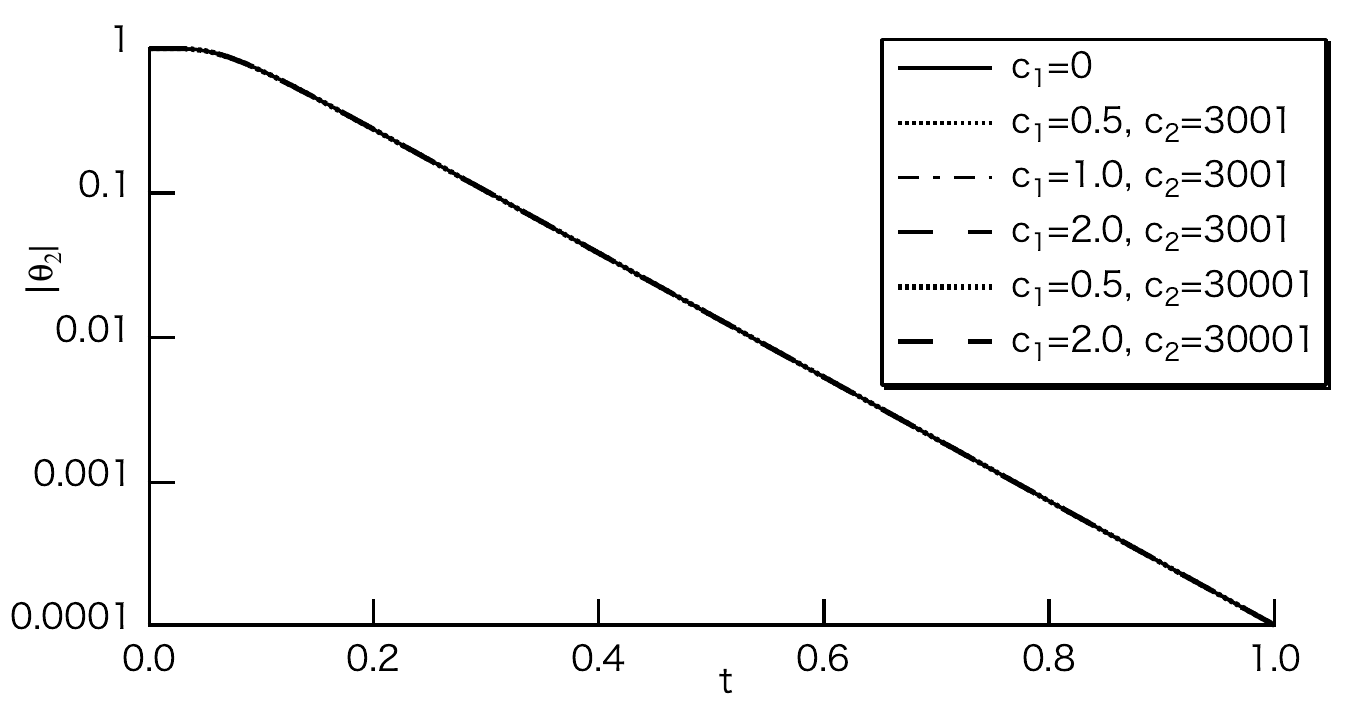}

\caption{The absolute temperature variation at point 2 for the power-law-type
thermal diffusivity. For any cases, the variations are almost identical.\label{fig3}}
\end{figure}

Figure \ref{fig3} shows the absolute temperature variation at point
2 for the case of power-law-type thermal diffusivity. The variation
is independent of the coefficients $c_{1}$ and $c_{2}$, and its
exponential decay rate 9.867 agrees well with that of (100) mode.
Please note that the initial heat input exclusively excites ($n$00)
($n$: integer) modes and that (100) is the SDM among them. Herein,
it is important that the decay rate of the SDM are not affected by
the temperature-dependent thermal diffusivity.

\begin{figure}
\includegraphics[scale=0.7]{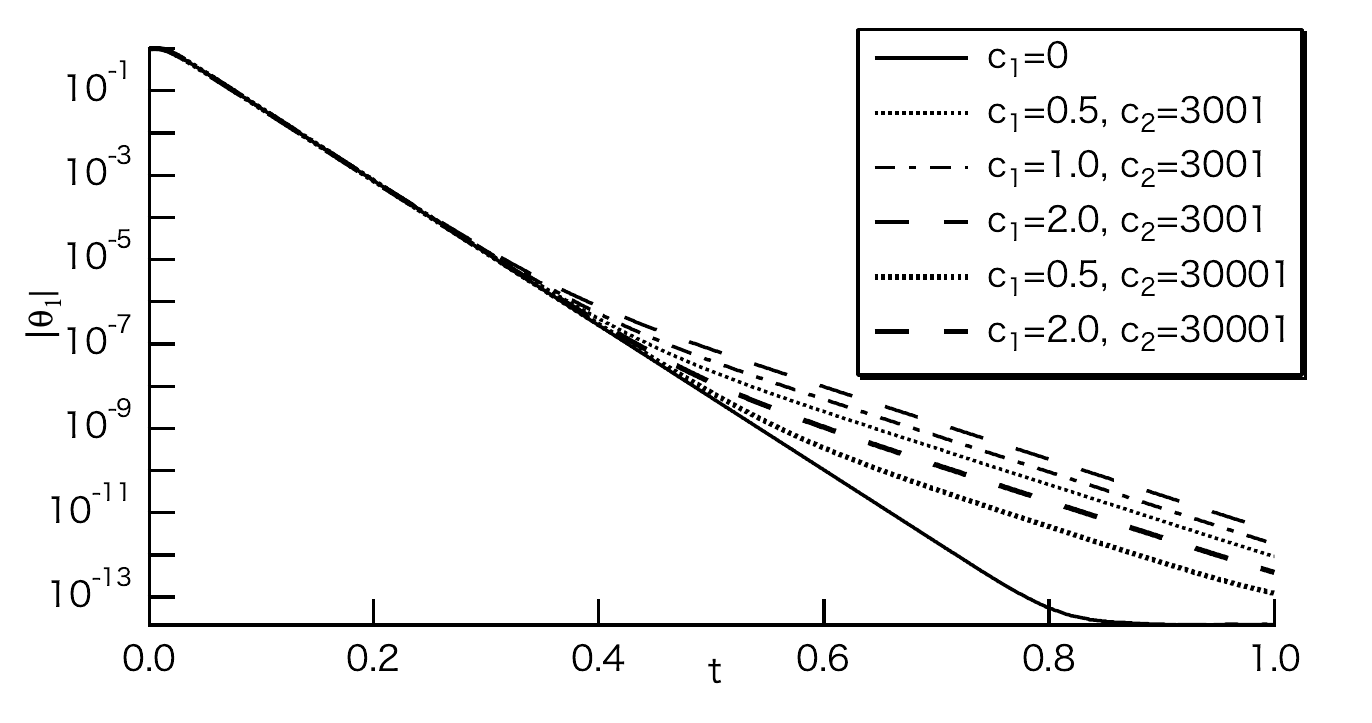}

\caption{The absolute temperature variation at point 1 for the power-law-type
thermal diffusivity. Thin and thick lines indicate the cases of $c_{2}=3001$
and 30001, respectively. Initially all lines are along the case of
$c_{1}=0$, i.e. the case of constant diffusivity. Eventually, the
lines separate and show the dependence on $c_{1}$ and $c_{2}$.\label{fig4}}
\end{figure}

In contrast, as shown in Fig. \ref{fig4}, the temperature variation
at point 1, bends at around $t=0.3\sim0.4$ when $c_{1}$ is nonzero
(not to be confused with a plateau at around $t=0.8$ for the case
of a constant thermal diffusivity, $c_{1}=0$, caused by the round-off
error around $10^{-13}$ ). The initial exponential decay ($t\lesssim0.3$)
does not depend on $c_{1}$ and $c_{2}$, and its decay rate 39.47
is in accord with that of (200) mode: the rSDM detected at point 1
is (200) mode because both excitation and detection are made on antinodal
region of (200) mode. The (100) mode can not be observed on the point
1 because the point is the node of (100) mode. However, the later
exponential decay depends on $c_{1}$ and $c_{2}$. For both of the
two $c_{2}$ cases, the decay rate ranges from 19.27 $(c_{1}=0.5,\textrm{ }c_{2}=30001)$
to 19.89 $(c_{1}=0.5,\textrm{ }c_{2}=3001)$ and shows good agreement
with $2\lambda_{100}(=19.74)$. This is the correction caused by the
dependency, observed on the (200) mode. Please note, herein, that
$\hat{v}_{100}^{2}$ corresponds to $\hat{v}_{200}$ except its constant
component. 

As $c_{2}$ is raised from 3001 to 30001, the bent point is delayed
from $t$=0.3 to 0.4: the temperature-dependent property is weakened.
Along the change, $|a_{1}|$ decreases approximately to 1/10. It implies
that the parameter $a_{1}$ is significant for quantifying the dependent
property (see Appendix \ref{sec:depnd_ind}).

\section{DISCUSSIONS AND CONCLUSIONS}

The reduction of the Biot number $Bi$, i.e. sufficient insulation
on the surface of specimen, holds the key to applicability of TMS.
There are three reasons for that. First, we can not a priori know
the distribution of heat transfer rate. We can define thermal modes
for not the Neumann but the Robin boundary condition as explained
in Sec. \ref{sec:basic_formulation}. Such modes, however, highly
depend on the fined distribution of heat transfer rate and, therefore,
it is almost impossible for us in advance to know nodal and antinodal
points for specifying a thermal mode. Second, we can not observe truly
slowest decaying mode, i.e. uniform steady mode of zero decay rate.
We have no choice but to regard a measured temperature decay rate
as a non-zero decay rate of a target mode. If the uniform steady mode
changes its eigenvalue to $S_{e}Bi/V$ by the minute heat transfer
on the surface, described in Sec. \ref{subsec:effects_h}, the measured
decay rate must be lower value by the contamination of the minute
decay rate of the uniform SDM and, therefore, the thermal diffusivity
is underestimated. Third, the decay rates of various modes are affected
themselves by the heat transfer rate. The corrections were obtained
by a perturbation analysis described in Sec. \ref{subsec:effects_h}.

In general, we can not regulate the corrections $\hat{\lambda}_{i}^{(n)}$
obtained in Sec. \ref{subsec:effects_h} because they depend on given
conditions such as the shape of specimen. All we can do is, therefore,
reduce the Biot number $Bi$ so that $S_{e}Bi/V$ would be far smaller
than the decay rate $\hat{\lambda}_{i}$ of a target mode. This is
the de-facto insulated condition. Since the number is defined by $Bi=hL/k$,
we can decrease the value by vacuuming or the reduction of typical
specimen size $L$. Or it becomes small when a given specimen has
high thermal conductivity. As explained in Sec. \ref{subsec:effects_h},
most measurement techniques request for the number to be sufficiently
large. That is the reason why it is difficult to apply the methods
to a small, like the order of 1mm \cite{ogi16}, high-conductivity
specimen. In contrast, TMS can rather conduct the measurement of such
specimens with high-precision. It is certain that this is the major
advantage of TMS method. 

However, we should not restrict the application of TMS to the small-size,
large thermal-diffusivity specimen. As described in Sec. \ref{subsec:temp-dependency},
the measurement of the decay rate of a specified mode has two advantages.
First, the decay rate is measured after time passes sufficiently.
In this final stage, the temperature variation within the specimen
is very small, e.g. less than 0.03K (Ref. \onlinecite{ogi16}). The
property allows us to identify the state in measurement with great
accuracy as the temperature and pressure at thermal equilibrium without
the introduction of such an effective temperature \cite{laserflush1961_org}.
In addition, the decay rate of the SDM is identical with the rate
at the equilibrium, not affected by the dependence of the thermal
conductivity on temperature, as explained in Sec. \ref{subsec:temp-dependency}.
In this sense, TMS truly overcomes the problem of the temperature
dependency. This is also the great advantages of TMS. It is meaningful,
therefore, to measure a low-diffusivity specimen in conjunction with
some vacuuming techniques. 

As explained in Sec. \ref{sec:intro}, the Kapitza resistive film,
i.e. the contact of a thin film with a specimen through the Kapitza
resistance, is inevitable for most measurement techniques of thermal
diffusivity. TMS has strength in this respect because the method is
not so influenced by the Kapitza resistance in the sense that the
first-order corrections to the decay rate caused by the film are independent
of the Kapitza conductance, as explained in Sec. \ref{subsec:effects_resistive-films}.
As long as the film thickness is sufficiently thin, therefore, there
is no need for TMS to evaluate the conductance. This is the great
advantage of TMS while most methods are required the evaluation to
eliminate or to consider the first-order contributions \cite{fdtr2009_org}.
The dimensionless film thickness $l$ is the ratio of a dimensional
film thickness $l^{*}$ to a typical length $L$ of specimen. It is
easy for us to reduce the thickness ratio less than $10^{-3}$ and,
consequently, the errors caused by the films. We \cite{ogi16} actually
achieved the ratio of $5.0\times10^{-4}$ even for millimeter size
specimens.

Herein, we should note that we can make thermal modes not affected
by the resistive films. In Eq. (\ref{eq:cors_temp_depend}), $\mathbf{x}^{T}\mathbf{Ny}$
and $\mathbf{x}^{T}\mathbf{Dy}$ present the integrated product of
functions $x,y$ and of their gradients on an attached-film surface
$S_{e}$, respectively. In TMS, such a surface is positioned at the
excitation or detection point, selected so that it is at an antinode
of a target mode, say $i$th mode, and also at a node of a mode, say
$k$th mode, to be eliminated. The mode-selection principle does remove
the first-order correction $w_{i,k}^{(1)}$ (Eq. (\ref{eq:c_i,k^(1)_tempd})),
significant for TMS.

This is also the case for the effects of heat transfer rate. If we
can insulate the surface of the specimen except the excitation and
detection points, similar discussion deduces that the first-order
correction $w_{i,k}^{(1)}$ (Eq. (\ref{eq:C_i,k^(1)_ht})) of thermal
mode vanishes. The condition is naturally fulfilled when we are to
insulate the surface of specimen as much as possible: such insulation
is impossible near the excitation and detection points. In both cases,
therefore, thermal modes are assumed to be kept unchanged for the
cases of small heat transfer rate and thin Kapitza resistive film.
It is significant for the mode-selection principle.
\begin{acknowledgments}
HI is grateful to Dr. G. Kawahara and Dr. M. Shimizu of Osaka University,
and Dr. E. Sasaki of Akita University for valuable comments and discussions.
This work is supported by JSPS KAKENHI Grant Number JP16K13719. 
\end{acknowledgments}

\appendix

\section{HEAT CONDUCTION EQUATION ON A SPECIMEN WITH KAPITZA RESISTIVE FILMS\label{sec:hceq_Kap}}

On the platform of the ODEs described in Sec. \ref{subsec:effects_h},
we shall incorporate the Kapitza resistive films as a boundary condition
within a portion of the specimen surface, and derive a closed heat
conduction equation for an arbitrary-shape specimen with arbitrary-shape
resistive films. 

Now we begin with a formulation of a film (domain $V'$) attached
to a portion $Se$ on a specimen. Application of the form (\ref{eq:unif_goveq_a})
to the film yields the evolution equation for film temperature $\bm{\uptheta'}$
as follows: \begin{subequations}\label{eq:goveq_rfilm}

\begin{equation}
\mathbf{M'}\frac{d\bm{\uptheta'}}{dt}=-\mathbf{\Gamma'}\bm{\uptheta'}+\mathbf{c'},\label{eq:goveq_vec_thetad}
\end{equation}
where

\begin{equation}
\mathbf{M'}=\{M'_{\alpha\beta}:\textrm{ }\mathit{M'_{\alpha\beta}=C'\int_{V'}\varphi_{\alpha}\varphi_{\beta}dV},\textrm{ }1\leq\alpha,\beta\leq N\},\label{eq:M'_alpbet}
\end{equation}

\begin{equation}
\mathbf{\Gamma'}=\{\varGamma'_{\alpha\beta}:\textrm{ }\mathit{\Gamma'_{\alpha\beta}=k'\int_{V'}\sum_{j}\frac{\partial\varphi_{\alpha}}{\partial x_{j}}\frac{\partial\varphi_{\beta}}{\partial x_{j}}dV},\textrm{ }1\leq\alpha,\beta\leq N\},\label{eq:Gam_alpbet}
\end{equation}

\begin{equation}
\mathbf{c'}=\{c'_{\alpha}:\textrm{ }\mathit{c'_{\alpha}=\int_{S_{e}}f_{n}\varphi_{\alpha}dS},\textrm{ }1\leq\alpha\leq N\},\label{eq:c'_alpha}
\end{equation}
\end{subequations}\noindent and the variable $C'$ is the ratio
of volumetric specific heat of the film to that of the specimen. The
heat conductivity ratio $k'$ is similarly defined. In this formulation
the flux $f_{n}$ is defined as the heat flux in the direction of
inward pointing normal vector on the film-specimen boundary $Se$:
by use of the thermal contact condition, it turns out to be the outward
normal heat flux from the specimen.

Expanding the flux $f_{n}$ by the basis functions $\varphi_{\beta}$
as follows

\[
\left.f_{n}(\mathbf{x},t)\right|_{\mathbf{x}\in S_{e}}=\left.\sum_{\beta}f{}_{\beta}(t)\varphi_{\beta}(\mathbf{x})\right|_{\mathbf{x}\in S_{e}},
\]
and substituting into Eq. (\ref{eq:c'_alpha}) we obtain\begin{subequations}

\begin{equation}
\mathbf{c'}=\mathbf{Nf},\label{eq:def_c'}
\end{equation}
where

\begin{equation}
\mathbf{f}\equiv(f_{1},\cdots,f_{N}),\label{eq:def_f}
\end{equation}

\begin{equation}
\mathbf{N}=\{N{}_{\alpha\beta}:\textrm{ }\mathit{N_{\alpha\beta}=\int_{Se}\varphi_{\alpha}\varphi_{\beta}dS},\textrm{ }1\leq\alpha,\beta\leq N\}.\label{eq:def_N}
\end{equation}

\end{subequations}

In order to obtain a solution to hold the quiescent initial condition,
the Laplace transform of Eq. (\ref{eq:goveq_vec_thetad}) with a complex
variable $s$ is utilized to find

\begin{equation}
\bar{\mathbf{f}}=\mathbf{\Xi'}\bm{\bar{\uptheta}'},\textrm{ }\mathbf{\Xi'}\equiv\mathbf{N}^{-1}(s\mathbf{M'}+\mathbf{\Gamma'}),\label{eq:tranf_flux}
\end{equation}
where $\bar{\mathbf{f}}$ and $\bm{\bar{\uptheta}'}$ are the Laplace
transforms of $\mathbf{f}$ and $\bm{\uptheta'}$, respectively. Herein,
as defined by Eq. (\ref{eq:def_N}), $\mathbf{N}$ is positive-definite
and, therefore, its inverse can be defined.

The Kapitza conductance $\alpha$ at the film-specimen interface makes
a surface temperature gap on both sides. When $\bm{\bar{\uptheta}}$
denote the Laplace transform of the temperature $\bm{\uptheta}$ on
the specimen surface $Se$, a dimensionless thermal contact condition
leads to $\bar{\mathbf{f}}=l_{K}^{-1}(\bm{\bar{\uptheta}}-\bm{\bar{\uptheta}'})$,
where $l_{K}\equiv(k/\alpha)/L$ is the ratio of the so-called Kapitza
length $k/\alpha$ to the typical length. Substituting the relation
into Eq. (\ref{eq:tranf_flux}) yields\begin{subequations}\label{eq:def_As}

\begin{align}
\bar{\mathbf{f}} & =(\mathbf{I}+l_{K}\mathbf{\Xi'})^{-1}\mathbf{\Xi'}\bm{\bar{\uptheta}}\nonumber \\
 & =(\mathbf{I}-l_{K}\mathbf{\Xi'}+l_{K}^{2}\mathbf{\Xi'}^{2}-l_{K}^{3}\mathbf{\Xi'}^{3}+\cdots)\mathbf{\Xi'}\bm{\bar{\uptheta}}\nonumber \\
 & \equiv\mathbf{A}'\bm{\bar{\uptheta}}\\
 & \equiv(\mathbf{A}'_{0}-s\mathbf{A}'_{1}+\cdots+(-s)^{n}\mathbf{A}'_{n}+\cdots)\bm{\bar{\uptheta}}.\label{eq:rel_f_and_theta}
\end{align}
\end{subequations}

On the other hand, the evolution equation (\ref{eq:goveq_vec_theta})
of specimen temperature is modified to be

\begin{equation}
\mathbf{M}\frac{d\bm{\uptheta}}{dt}=-\mathbf{\Gamma}\bm{\uptheta}-\mathbf{Nf}.\label{eq:goveq_specim}
\end{equation}
Substituting of the inverse Laplace transform of Eq. (\ref{eq:rel_f_and_theta})
into the above equation, we eventually obtain a closed governing equation
for specimen temperature as follows:

\begin{equation}
\mathbf{M}\dot{\bm{\uptheta}}+\mathbf{\Gamma}\bm{\uptheta}+\mathbf{N}(\mathbf{A}'_{0}\bm{\uptheta}-\mathbf{A}'_{1}\dot{\bm{\uptheta}}+\mathbf{A}'_{2}\ddot{\bm{\uptheta}}+\cdots+(-1)^{n}\mathbf{A}'_{n}\bm{\uptheta}^{(n)}+\cdots)=0.\label{eq:finaleq_theta}
\end{equation}

If we have more than one ($R$) resistive films, the third term of
the left-hand side of the equation should be replaced by 

\[
\sum_{i=1}^{R}\mathbf{N}_{i}(\mathbf{A}_{0}^{(i)'}\bm{\uptheta}-\mathbf{A}_{1}^{(i)'}\dot{\bm{\uptheta}}+\mathbf{A}_{2}^{(i)'}\ddot{\bm{\uptheta}}+\cdots).
\]

\section{INDICATOR OF DEPENDENCY OF THERMAL DIFFUSIVITY ON TEMPERATURE\label{sec:depnd_ind}}

In order to elucidate the importance of parameter $a_{1}$ as the
indicator of temperature-dependent properties, in this section, two
types of thermal diffusivity are tested:

\[
a(\theta)=\begin{cases}
e^{-c_{3}\theta} & ;\textrm{ exponential type }(a_{1}=-c_{3}),\\
\frac{1}{\left(1+\theta^{2}\right)^{c_{4}}} & ;\textrm{ modified power-law type }(a_{1}=0),
\end{cases}
\]
where coefficients $c_{i}\textrm{ }(i=3,4)$ are also supposed to
be positive number. 

The coefficient $a_{1}$ for each case is obtained from Eq. (\ref{eq:expand_a}).
By use of dimensional thermal diffusivity $a^{*}(T)$, it can be rewritten
as

\begin{equation}
a_{1}=\left.\frac{1}{a}\frac{\partial a}{\partial\theta}\right|_{\theta=0}=\frac{\Delta T}{a^{*}(Te)}\left.\frac{\partial a^{*}}{\partial T}\right|_{T=T_{e}},\label{eq:dimensional_a1}
\end{equation}
and, therefore, it can be regarded as the normalized exponential increase
rate of thermal diffusivity at the steady state.

Most pure metals \cite{aiphandb,jsmehandb} have the dimensional coefficient
$c_{3}^{*}$ of the order of $10^{-4}\sim10^{-3}$, and have the coefficient
$c_{1}<1$. If the temperature rise $\Delta T$ is of the order of
$0.01\sim0.1$K (see, for example, Ref. \onlinecite{ogi16}), the
coefficient $c_{3}(=c_{3}^{*}\Delta T)$ is estimated to be $10^{-6}\sim10^{-4}$.
On the other hand, the coefficient $c_{2}\equiv T_{e}/\Delta T$ is
of the order of $10^{3}\sim10^{4}$ when $T_{e}$ is near a room temperature
and $\Delta T$ is $0.01\sim0.1$K. In this study $c_{2}$ is fixed
at 3001 or 30001, and $c_{1}$ is changed from 0 to 2.0. It follows
that the order of $c_{3}(\sim c_{1}/c_{2})$ is less than that of
$10^{-4}\sim10^{-3}$. $c_{4}$ is changed from 0 to 1.0 because double
the value corresponds to $c_{1}$.

The physical model and the discretization of the heat conduction equation
are the same as those explained in Sec. \ref{subsec:simulation}.
Most computations (except one case mentioned below) were performed
by double precision.

First of all, the computations by the exponential-type thermal diffusivity,
corresponding to Figs. \ref{fig3} and \ref{fig4}, are performed
under the same initial condition. With the coefficient $c_{3}$ held
fixed at $c_{1}/c_{2}$, i.e. with the corresponding two $a_{1}$'s
being identical, we can confirm that temperature variations for the
exponential type agree up to the first five digit numbers with the
variations of the power-law-type diffusivity, and they can not be
distinguished on Figs. \ref{fig3} and \ref{fig4}. These results
indicate that the coefficient $a_{1}$ is important for the temperature-dependent
properties.

\begin{figure}
\includegraphics[scale=0.7]{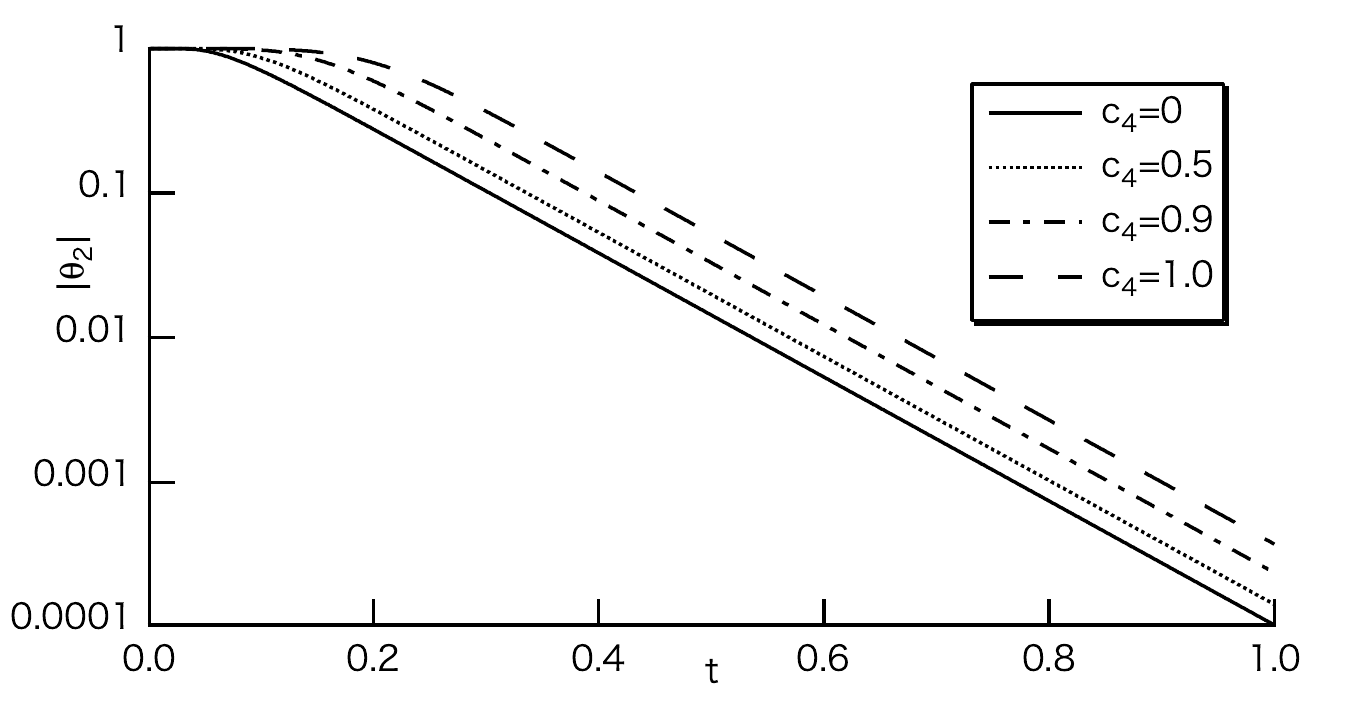}

\caption{The absolute temperature variation at point 2 for the modified power-law-type
thermal diffusivity. Unlike Fig. \ref{fig3}, the initial variations
are strongly affected by $C_{4}$. In the final stage, however, every
line decays with the same rate.\label{fig5} }
\end{figure}

When we choose the modified power-law-type diffusivity, temperature
variations become simple. In this type of dependency $a_{1}$ is fixed
at zero. Figure \ref{fig5} shows the temperature variation at point
2. Since the diffusivity is far smaller than that of power-law type
for $\theta>10$, the temperature initially remains constant when
$c_{4}$ is large. But it eventually shows exponential decay whose
decay rate ranges from 9.864 ($c_{4}$=0.9) to 9.867 ($c_{4}$=0),
in agreement with that of (100) mode. 

\begin{figure}
\includegraphics[scale=0.4]{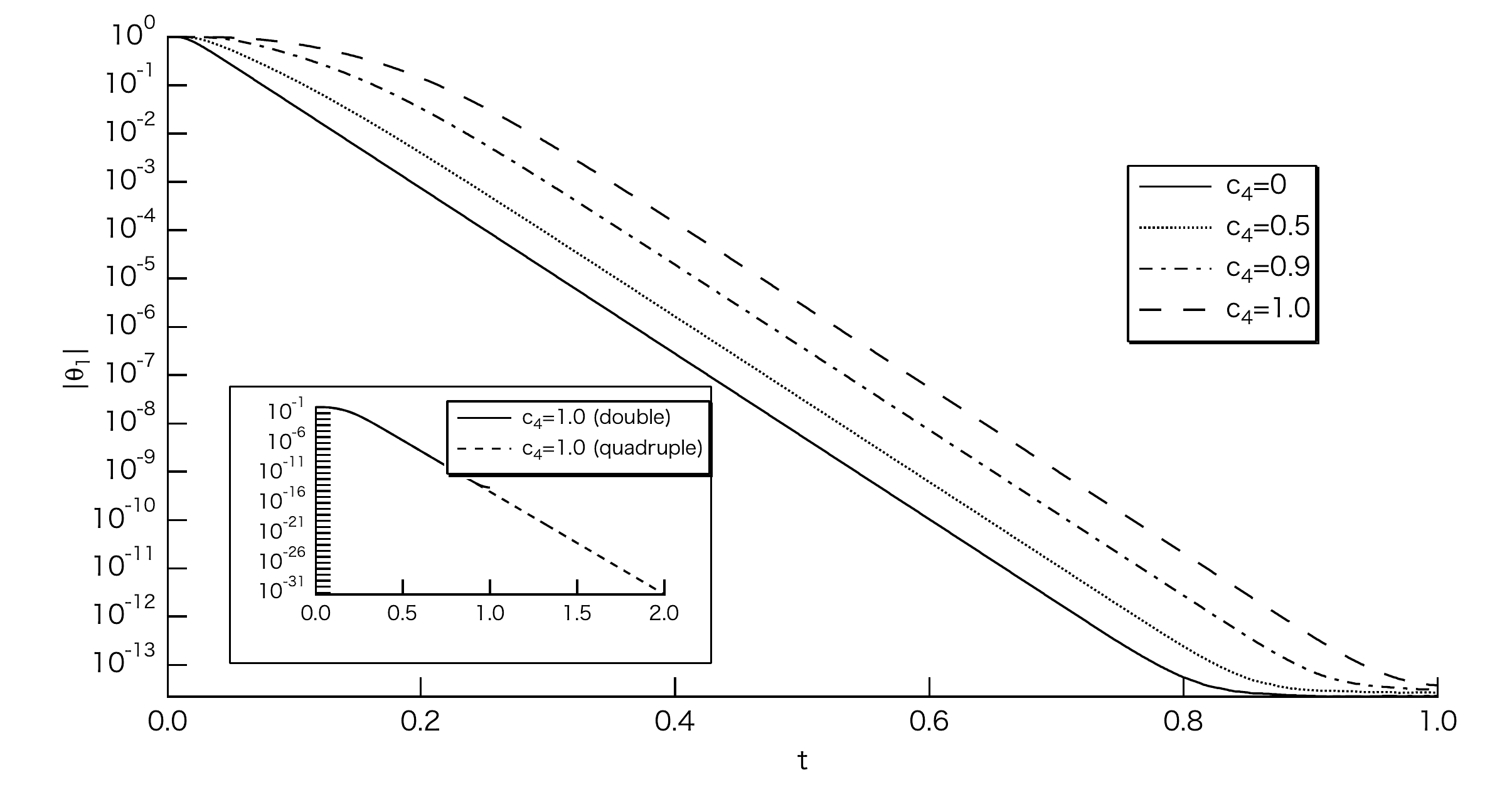}

\caption{The absolute temperature variation at point 1 for the modified power-law-type
thermal diffusivity. Unlike Fig. \ref{fig4}, every line decays with
a single rate, saturating around the order of its round-off error,
$10^{-13}$. On the other hand, the inset shows the result of quadruple-precision
computation. The line decays to reach around the order of its round-off
error, $10^{-31}$. Even with this accurate computation, we can not
find the effects of $a_{2}$ or higher-order corrections.\label{fig6}}
\end{figure}

As shown in Fig. \ref{fig6}, the variation at point 1 does not show
the bent: its exponential decay rate ranges from 39.32 ($c_{4}$=0.9)
to 39.47 ($c_{4}$=0), almost identical with that of (200) mode. The
effects of higher-order corrections or $a_{2}$ can appear in the
final, minute temperature variation in the vicinity of the thermal
equilibrium. In this study, therefore, a computation by quadruple
precision is performed and the result is shown in the inset of Fig.
\ref{fig6}. The temperature shows a single exponential decay eventually
to reach the order of round-off error $10^{-31}$, lower than the
limit of temperature measurement. There is no evidence that higher-order
corrections appear. The results clarified that the normalized exponential
increasing rate $a_{1}$ of thermal diffusivity can be regarded as
the indicator to quantify the effects of temperature-dependent thermal
diffusivity.

From the expression (\ref{eq:dimensional_a1}) of $a_{1}$, we might
expect that it is reducible by diminishing the (typical) temperature
rise $\Delta T$ caused by the heat input in measurement. However,
it depends on the property of a specimen. From the discussion in Fig.
\ref{fig4}, we confirmed that the bent point $t_{b}$, i.e. the time
when the effects of temperature dependence appear, is delayed from
0.3 to 0.4. We can extrapolate the result to estimate that $a_{1}$
should be less than $1.0\times10^{-8}$ to hold the condition that
$t_{b}=1$ for the accurate measurement of the decay rate. For most
pure metals the dimensional exponential increase rates of diffusivity
are typically the order of $10^{-4}\sim10^{-3}$ (1/K), it follows
that $\Delta T$ must be less than $10^{-5}\sim10^{-4}$ (K). Thereby
it is difficult for us accurately to detect temperature evolution.

Thus, we need the plan to avoid such a situation. The temperature
change shown in Fig. \ref{fig4} is caused by the selection of not
the SDM {[}(100) mode{]} but (200) mode (rSDM) as the target among
excited modes. If the point 2 were chosen, we could have measured
the decay rate $\lambda_{SDM}$ of the SDM as shown in Fig. \ref{fig3}.
It follows that the detection mode must be in accord with the SDM
among the excited modes except the case that we measure in a positive
manner the decay rate of $2\lambda_{SDM}$ for the case of temperature-dependent
thermal diffusivity.

\bibliographystyle{apsrev4-1}
\bibliography{tms}

\end{document}